\documentstyle[12pt]{article}

\makeatletter
\def\vereq#1#2{\lower3pt\vbox{\baselineskip1.5pt \lineskip1.5pt
\ialign{$\m@th#1\hfill##\hfil$\crcr#2\crcr\sim\crcr}}}
\makeatother

\def\lesssim{\mathrel{\mathpalette\vereq<}}

\makeatletter
\def\lesssim{\mathrel{\mathpalette\vereq<}}
\def\vereq#1#2{\lower3pt\vbox{\baselineskip1.5pt \lineskip1.5pt
\ialign{$\m@th#1\hfill##\hfil$\crcr#2\crcr\sim\crcr}}}

\makeatother

\newcommand{\vev}[1]{$\langle #1 \rangle$}
\newcommand{\beq}{\begin{equation}}
\newcommand{\eeq}{\end{equation}}
\newcommand{\bfbar}[1]{$\overline{{\rm {\bf #1}}}$}

\newcommand{\remove}[1]{}

\begin{document}
\begin{titlepage}
\begin{center}
\today     \hfill    LBL-37810\\
{}~{} \hfill UCB-PTH-95/32\\

\vskip .25in

{\large \bf Phenomenology of Minimal SU(5) Unification\\
with Dynamical Supersymmetry Breaking}
%
\footnote{This work was supported in part by the Director, Office of
Energy Research, Office of High Energy and Nuclear Physics, Division of
High Energy Physics of the U.S. Department of Energy under Contract
DE-AC03-76SF00098 and in part by the National Science Foundation under
grant PHY-90-21139.}

\vskip 0.3in

Christopher D. Carone$^1$ and Hitoshi Murayama$^{1,2}$

\vskip 0.1in

{{}$^1$ \em Theoretical Physics Group\\
     Lawrence Berkeley Laboratory\\
     University of California, Berkeley, California 94720}

\vskip 0.1in

{{}$^2$ \em Department of Physics\\
     University of California, Berkeley, California 94720}

\end{center}

\vskip .3in

\begin{abstract}
We consider the constraints from proton decay and $b$-$\tau$
unification in the minimal supersymmetric SU(5)
grand unified theory with a `visible' dynamical
supersymmetry breaking sector.  We show how the presence of
vector-like messenger fields and the constrained superparticle mass
spectrum affect the phenomenology of the model.  We include the
messenger fields in our renormalization group analysis between the
messenger scale ($\sim 100$ TeV) and the GUT scale.  We show that the
simplest model of this type, a minimal SU(5) GUT with an additional
{\bf 5}+\bfbar{5} of messenger fields is excluded by
the constraints from proton decay and $b$-$\tau$ unification.
\end{abstract}

\end{titlepage}
\renewcommand{\thepage}{\roman{page}}
\setcounter{page}{2}
\mbox{ }

\vskip 1in

\begin{center}
{\bf Disclaimer}
\end{center}

\vskip .2in

\begin{scriptsize}
\begin{quotation}
This document was prepared as an account of work sponsored by the United
States Government. While this document is believed to contain correct
information, neither the United States Government nor any agency
thereof, nor The Regents of the University of California, nor any of their
employees, makes any warranty, express or implied, or assumes any legal
liability or responsibility for the accuracy, completeness, or usefulness
of any information, apparatus, product, or process disclosed, or represents
that its use would not infringe privately owned rights.  Reference herein
to any specific commercial products process, or service by its trade name,
trademark, manufacturer, or otherwise, does not necessarily constitute or
imply its endorsement, recommendation, or favoring by the United States
Government or any agency thereof, or The Regents of the University of
California.  The views and opinions of authors expressed herein do not
necessarily state or reflect those of the United States Government or any
agency thereof, or The Regents of the University of California.
\end{quotation}
\end{scriptsize}

\vskip 2in

\begin{center}
\begin{small}
{\it Lawrence Berkeley Laboratory is an equal opportunity employer.}
\end{small}
\end{center}

\newpage
\renewcommand{\thepage}{\arabic{page}}
\setcounter{page}{1}
\section{Introduction}\label{sec:intro}

Supersymmetric grand unification is one of the viable possibilities
for the physics that lies beyond the standard model \cite{sur}.  Interest
in supersymmetric grand unified theories (SUSY GUTs) has been motivated
primarily by two observations: ({\em i}) supersymmetry eliminates the
quadratic divergences in radiative corrections to the Higgs mass, which
can destabilize the hierarchy between the GUT and electroweak
scales \cite{msr}, and ({\em ii}) gauge coupling unification can be
achieved to considerable accuracy provided that superparticle masses
are $\lesssim 1$ TeV \cite{ymsr}.  While this picture may turn out to
be correct, it provides no explanation for why the supersymmetry breaking
scale is so low (for example, in comparison to the Planck scale).  In
addition, it gives no explanation for the near degeneracy of the squark
masses, necessary to avoid large flavor-changing neutral current (FCNC)
effects.  While supersymmetry can resolve ({\em i}) and ({\em ii}) above,
the origin, scale and pattern of supersymmetry breaking masses required
to produce a viable phenomenology remains a separate and important puzzle.

Recently proposed models of dynamical supersymmetry breaking (DSB) provide
possible solutions to these remaining problems  \cite{nd1,nd2}.  If
supersymmetry is broken by nonperturbative dynamics triggered when some
asymptotically-free gauge coupling $g(\Lambda)$ becomes large, then we
would expect
\beq
\Lambda \approx m_0 \exp \frac{8\pi^2}{b g^2(m_0)}
\eeq
where $b<0$ is a beta function, and $m_0$ is some high scale, like
$M_{\rm Planck}$. The exponential suppression can account for a hierarchy
between the SUSY-breaking scale and the Planck scale.  Furthermore, the
models that have been proposed so far employ a mechanism \cite{wise} by
which supersymmetry breaking is transmitted to the ordinary particles
through loop diagrams involving vector-like ``messenger" fields, that carry
electroweak quantum numbers, and ordinary gauge interactions.  When the
messenger fields feel SUSY breaking originating from the dynamical SUSY
breaking sector of the theory, they transmit it to the ordinary squarks
via these diagrams, which are flavor independent \cite{nd1,nd2}.  As as
a result, the squarks of different generations remain degenerate, and
the FCNC problem is naturally avoided.

In this paper, we will comment on the GUT phenomenology of models
of this type.  In the present context, GUT refers to the unification
of the ordinary gauge groups of the standard model, but not to
the unification of these groups with the additional gauge groups
responsible for dynamical supersymmetry breaking.
We restrict our discussion to the minimal SU(5) grand unified
model \cite{DGS} for definiteness.  We will first argue that
the existence of vector-like multiplets with electroweak quantum numbers
at a scale $\sim 100$ TeV is generic to any workable model.  We will
catalog the possible particle content of the messenger sector that is
allowed by the requirement of perturbative gauge coupling unification,
and we state the predicted squark and gaugino mass relations that follow
in each case.  Using this information, and including the messenger
fields in our renormalization group analysis, we determine the bounds
from the nonobservation of proton-decay, and from the requirement
of $b$-$\tau$ Yukawa unification.  We conclude that a minimal SU(5) GUT
with the simplest messenger sector possible is excluded by the lower
bounds on the proton half-life.

\section{Messenger Sector} \label{sec:msec}

The part of the messenger sector that is relevant to our analysis
involves those fields which carry electroweak quantum numbers.  These
fields transmit SUSY breaking to the ordinary sector via loop diagrams
involving electroweak gauge interactions.  Since these diagrams have
appeared in a number of places in the literature \cite{nd1,nd2,wise},
we do not display them again here.  In the model of Ref.~\cite{nd2},
the relevant part of the messenger sector superpotential is
\beq
W_m = \lambda_D S\, \overline{D}\, D + \lambda_l S\,\overline{l}\, l
\label{eq:msp}
\eeq
where the fields have the quantum numbers $D \sim (3,1)_{-1/3}$
$\overline{D} \sim (\overline{3},1)_{1/3}$, $l \sim (1,2)_{-1/2}$,
and $\overline{l} \sim (1,2)_{1/2}$ under the standard model gauge group,
and $S$ is a singlet chiral superfield.  This particle content forms
full SU(5) multiplets, ${\bf 5} + \bar{\bf 5}$, so that the apparent gauge
unification at $\sim 2 \times 10^{16}$~GeV is preserved.  The vacuum
expectation value (vev) of the scalar component of $S$, \vev{S}, determines
the SU(3)$\times$SU(2)$\times$U(1) invariant masses in (\ref{eq:msp}),
while the vev of the $F$ component of $S$, \vev{F_S}, parametrizes the
degree of SUSY breaking. In Ref.~\cite{nd2} the 3-2 model \cite{ads} is
assumed as the source of DSB, and the authors show that the remaining
portion of the messenger superpotential can be constructed so that DSB
in the 3-2 sector generates vevs for both $S$ and $F_S$.

We first would like to argue that the portion of the messenger
superpotential given in (\ref{eq:msp}) is generic to a wide variety
of realistic models in which SUSY breaking is transmitted to the
ordinary sector via gauge interactions.  If the messenger quarks and
leptons are vector-like under the standard model gauge group, and we
allow no dimensionful couplings in the superpotential, then the couplings
in (\ref{eq:msp}) will be present. The latter requirement is a
philosophical one, namely, that all mass scales in the theory be
generated via dimensional transmutation.  One might imagine constructing a
model in which the messenger fields are chiral rather than vector-like under
the standard model gauge group.  The general problem of a model of this type is
that radiatively-generated gaugino masses are too small.  The one-loop diagram
responsible for generating a gaugino mass necessarily involves chirality
flips on the fermion and scalar lines.  These chirality flips are
proportional to the messenger fermion masses, $m_f$, and therefore the
gaugino masses are of order $(\alpha/4\pi) m_f^2/m_{SUSY}$. Since $m_f$ is of
order the weak scale, the gauginos in this scenario are unacceptably
light.\footnote{We do not consider the possibility of light gluinos in this
paper \cite{gfarrar}.}

One might also imagine that the
messenger quarks and leptons are vector-like under the standard model
gauge group, but carry nonstandard quantum numbers as well.  In this
case, it is likely that the additional gauge couplings
would be perturbative.  The only gauge group that is nonperturbative
is the one in the DSB sector, and introducing additional particles that
transform under it could lead to disaster in two ways: the vacuum
structure of the theory may change so that supersymmetry is restored,
or the multiplicity of messenger quarks and leptons may be too large to
retain perturbative unification of the ordinary gauge couplings \cite{nd1}.
We know of no workable model in which the messenger quarks and leptons
couple directly to a strongly interacting group from the DSB sector.
If the messenger quarks and leptons couple to a nonstandard gauge group that is
perturbative (like SU(3) in Ref.~\cite{nd1}), one might still worry that the
conclusions of the two-loop proton decay analysis that we will present in the
next section could be altered. While the standard model gauge couplings
will run differently in this case, one can show that the quantities relevant
to our analysis (e.g., the mismatch of gauge couplings at the GUT scale,
mass ratios of particles at the messenger scale, etc.) will remain
unaltered and our conclusions will remain the same.

In what follows we assume minimal SU(5) unification, so that the
gauge structure of the theory is SU(5)$\times G_{{\rm DSB}}$.  In
$G_{{\rm DSB}}$ we include any nonstandard gauge groups that may
be necessary for communicating SUSY breaking to the full messenger
sector superpotential (e.g., the messenger hypercharge group discussed
in \cite{nd2}).  SU(5) gauge invariance implies that the messenger quarks
and leptons form complete SU(5) representations.  In the minimal case of
a {\bf 5}+\bfbar{5} in the messenger sector, the messenger superpotential
at the GUT scale has the form
\beq
W_m = \lambda \,S \overline{{\rm{\bf 5}}}\,{\bf 5} \,\, .
\eeq
However, below the GUT scale, SU(5) is broken, and we recover
the superpotential given in (\ref{eq:msp}).  The assumption of
unification allows us to compute $\lambda_D$ and $\lambda_l$
in terms of $\lambda$, by running these couplings down to the
messenger scale.  Once the messenger scale has been specified,
threshold corrections at this scale are calculable, and can be
included in our renormalization group analysis without introducing
any additional uncertainty.  This is true for representations larger
than {\bf 5}+\bfbar{5} as well.

Next, we must specify what other SU(5) representations are
allowed in the messenger sector.   Introducing additional SU(5)
multiplets preserves gauge unification, but the gauge coupling
at the GUT scale $\alpha_5(m_{GUT})$ increases as we add additional
multiplets.  If we require that $\alpha_5(m_{GUT})$ remains
perturbative, then we may add 1, 2, 3 or 4 ({\bf 5}+\bfbar{5}) pairs,
or a single ({\bf 10}+\bfbar{10}) pair, or
({\bf 5}+\bfbar{5})+({\bf 10}+\bfbar{10})
to the particle content
of the minimal SU(5) GUT.  Additional {\bf 5}s or {\bf 10}s,
or larger SU(5) representations will render $\alpha_5(m_{GUT})$
nonperturbative \cite{MMY}.

The messenger sector fields not only affect the renormalization
group analysis between the messenger scale $\Lambda$ and the
GUT scale (including the calculable threshold corrections
at the scale $\Lambda$) but also constrain the threshold
corrections at the weak scale.  As presented in
Ref.~\cite{nd2}, the radiatively-generated gaugino and
squark masses, $m_i$ and $\tilde{m}$, assuming a {\bf 5}+\bfbar{5} in
the messenger sector, are given by
\beq
m_{i}=\frac{g_i^2}{16 \pi^2} \frac{\langle F_S\rangle}
{\langle S\rangle} \,\, ,
\label{eq:mgaugino}
\eeq
\beq
\tilde{m}^2 = \sum_a 2 C_F^{(a)} \left( \frac{g^{(a)2}}
{16 \pi^2}\right)^2 \frac{\langle F_S \rangle^2}
{\langle S\rangle^2} \,\, ,
\label{eq:msquark}
\eeq
where $C_F$ is 3/4 for SU(2) doublets, 4/3 for SU(3) triplets, and
(3/5)$Y^2$ for U(1).  The quantity \vev{F_S}$/$\vev{S}$\equiv m_0$ has
dimensions of mass, and is of the same order as the messenger
scale $\Lambda$.  In a specific model, $m_0$ is calculable by minimizing
the messenger sector potential.  Once $m_0$ is fixed, the particle content
of the messenger sector completely determines the gaugino and squark
masses at the scale $\Lambda$; these can subsequently be run down to the
weak scale.  We explain how we fix the precise value of $m_0$
in the following section.  If the messenger sector consists
of $n_5$ {\bf 5}+\bfbar{5} pairs, then (\ref{eq:mgaugino}) and
(\ref{eq:msquark}) both scale as $n_5$, while the
ratio $m_{i}/\tilde{m}$ scales as $\sqrt{n_5}$.  In the case
of a {\bf 10}+\bfbar{10} pair, we obtain the same result for the
gaugino and squark masses as the case where $n_5=3$.  These observation
will be useful in our discussion of the proton decay bound in the
following section.

\section{Proton Decay Analysis} \label{sec:pdecay}

Our analysis of the proton decay constraints is similar
in spirit to that of Refs.~\cite{hmm,HMTY}.  By including
threshold corrections at the GUT scale, we can determine
the largest color-triplet Higgs mass that is consistent
with gauge coupling unification.  We can then constrain the
remaining free parameters involved in the proton decay
matrix element using the current lower bounds on the proton
half-life. Our algorithm is as follows:

{\em i}.  We fix $n_5$ or $n_{10}$, the number of {\bf 5}+\bfbar{5}
or {\bf 10}+\bfbar{10} pairs in the messenger sector.  The messenger
sector scale is set at 100 TeV, to insure that the superparticle masses
are of order the weak scale (recall equations (\ref{eq:mgaugino})
and (\ref{eq:msquark}) above).

{\em ii}. We define the GUT scale $M_{GUT}$ as the scale where
the SU(2) and U(1) gauge couplings unify.  We use the
input values $\alpha_1^{-1}(m_Z) = 58.96 \pm 0.05$ and
$\alpha_2^{-1}(m_Z) = 29.63 \pm 0.05$, that follow from
Ref.~\cite{ip}. We first run them up to $m_{top} = 176$~GeV using the
standard model renormalization group equations (RGEs).
We then numerically solve the two-loop RGEs
with the supersymmetric particle content, taking into
account the change in the one- and two-loop beta functions as we cross the
messenger scale.  The RGEs  and beta functions are provided in the appendix.

{\em iii}. Given an input value of $\alpha_3(m_Z)$, we
determine the mismatch between $\alpha^{-1}_3(M_{GUT})$, and
$\alpha_5^{-1}(M_{GUT})$, the GUT coupling determined in
step {\em ii}.  We ascribe this mismatch to the sum effects
of threshold corrections at the weak scale, the messenger scale,
and the GUT scale.  Using the one-loop expressions for
threshold corrections given in Ref.~\cite{hmm}, we find
that the mismatch
$\Delta \alpha^{-1}_3 \equiv \alpha^{-1}_3(M_{GUT}) -
\alpha_5^{-1}(M_{GUT})$ is given by
\beq
\Delta \alpha_3^{-1} = -\frac{1}{4\pi} \left[
\frac{12}{5} \ln \frac{M_{H_c}}{M_{GUT}}
+4 \ln \frac{m_{\tilde{g}}}{m_{\tilde{w}}}
-\frac{8}{5} \ln \frac{m_{\tilde{h}}}{m_{top}}
+\frac{12}{5} n_5 \ln \frac{m_D}{m_L} \right] \, .
\label{eq:tc}
\eeq
The first term gives the threshold correction at the GUT scale
as a function of the color-triplet Higgs mass $M_{H_c}$.
The next two terms give the largest \cite{hmm} threshold corrections
at the weak scale, depending on the gluino, wino, and higgsino
masses.  We checked the threshold corrections from scalars can be safely
neglected in this analysis.
The third term gives the threshold correction at
the messenger scale, from the splitting of the original $n_5$
{\bf 5}+\bfbar{5} pairs.  In the case of a {\bf 10}+\bfbar{10} pair,
we have a different particle content at the messenger scale,
and we should make the
replacements
\beq
\frac{12}{5}n_5 \log\frac{m_D}{m_L} \hspace{1cm}
\longrightarrow \hspace{1cm}
-\frac{6}{5} \log \frac{m_Q}{m_E}-\frac{18}{5} \log \frac{m_Q}{m_U}
\label{eq:tc10}
\eeq
where the new fields have the quantum number assignments
$Q\sim (3,2)_{1/6}$, $E\sim (1,1)_{-1}$ and $U\sim (3,1)_{-1/3}$.

Note that most of the variables in (\ref{eq:tc}) and (\ref{eq:tc10})
can be estimated reliably enough, making it possible to place an upper
bound on $M_{H_c}$.  We take $m_{\tilde{g}}/m_{\tilde{w}}=\alpha_3/
\alpha_2 \approx 3.5$, and set $m_{\tilde{h}}=1$ TeV to maximize $M_{H_c}$.
The ratio $m_Q/m_L$ can be computed by running the Yukawa couplings
in (\ref{eq:msp}) between the GUT scale and the messenger scale.  We use
a one-loop estimate of this ratio which takes into account the effect of
the gauge interactions on the running. We find
\beq
m_D/m_L \approx \left[\frac{\alpha_5^{-1}}{\alpha_1^{-1}(\Lambda)}
\right]^{\frac{1}{6}\frac{1}{\frac{33}{5}+n_5}} \,
\left[\frac{\alpha_5^{-1}}{\alpha_2^{-1}(\Lambda)}
\right]^{\frac{3}{2}\frac{1}{1+n_5}} \,
\left[\frac{\alpha_5^{-1}}{\alpha_3^{-1}(\Lambda)}\right]
^{-\frac{8}{3}\frac{1}{-3+n_5}}
\label{eq:mqml}
\eeq
for $n_5<3$.  For $n_5=3$ (the case where $\alpha_3$ no longer runs
at the one-loop level above the messenger scale), one makes the
substitution
\beq
\left[\frac{\alpha_5^{-1}}{\alpha_3^{-1}(\Lambda)}\right]
^{-\frac{8}{3}\frac{1}{-3+n_5}} \rightarrow
\left[\frac{\Lambda}{M_{GUT}}\right]^{-\frac{4}{3}
\frac{\alpha_3(\Lambda)}{\pi}}
\eeq
in eq.~(\ref{eq:mqml}).  Finally, the ratios that are
relevant in the {\bf 10}+\bfbar{10} case are given by
\beq
m_Q/m_E \approx \left[\frac{\alpha_5^{-1}}{\alpha_1^{-1}(\Lambda)}
\right]^{\frac{35}{288}}\,
\left[\frac{\alpha_5^{-1}}{\alpha_2^{-1}(\Lambda)}
\right]^{-\frac{3}{8}}\,
\left[\frac{\Lambda}{M_{GUT}}\right]^{-\frac{4}{3}
\frac{\alpha_3(\Lambda)}{\pi}} \, ,
\eeq
\beq
m_Q/m_U \approx \left[\frac{\alpha_5^{-1}}{\alpha_1^{-1}(\Lambda)}
\right]^{\frac{15}{288}}\,
\left[\frac{\alpha_5^{-1}}{\alpha_2^{-1}(\Lambda)}
\right]^{-\frac{3}{8}} \, .
\eeq

For any input value of $\alpha_3(m_Z)$, we determine the
mismatch $\Delta \alpha_3^{-1}$, and the maximum $M_{H_c}$ that follows
from (\ref{eq:tc}) or (\ref{eq:tc10}).  Note that in determining this upper
bound, we take into account the uncertainties in $M_{GUT}$
and $\alpha_5(M_{GUT})$ that follow from the experimental uncertainties
in $\alpha_1(m_Z)$ and $\alpha_2(m_Z)$ at 90\% confidence level. The results
are shown in Figure~1.   In the case where we ignore threshold corrections
at the messenger scale, Figure~1a, we see that the addition
of {\bf 5}+\bfbar{5} pairs tends to weaken the bound on $M_{H_c}$; in
the {\bf 10}+\bfbar{10} case the bound is strengthened.
However, we see that in the final result, Figure~1b, the upper bound
on $M_{H_c}$ is not much different from the minimal case.
This is due to an accidental cancellation between two
different effects, namely the two-loop contribution to the gauge coupling
evolution due to the additional fields, and the threshold correction at the
messenger scale.

{\em iv}.  Using our results shown in Figure~1, we can study the bounds
from the proton life-time.  Consider the mode
$n\rightarrow K^0 \overline{\nu}_\mu$. The nucleon life-time is
given in Ref.~\cite{hmm} as
\[
\tau(n\rightarrow K^0 \overline{\nu}_\mu) = 3.9\times 10^{31} \mbox{yrs.}
\,\,\times \,\,\,\,\,\,\,\,\,\,\,\,\,\,\,\,\,\,\,\,\,\,\,
\]
\beq
\left|\frac{0.003 {\rm GeV}^3}{\zeta}\frac{0.67}{A_S}
\frac{\sin 2\beta}{1+y^{tK}}\frac{M_{H_C}}{10^{17}{\rm GeV}}
\frac{{\rm TeV}^{-1}}{f(u,d)+f(u,e)} \right|^2  \, .
\label{eq:ptau}
\eeq
Here $\zeta$ parametrizes the uncertainty in the hadronic matrix
element, and is estimated to be between $0.003$ GeV$^3$ to $0.03$ GeV$^3$
\cite{ENR}.  We set $\zeta=0.003$ to be the most conservative.
$A_S$ represents the short distance renormalization of the proton-decay
operators, i.e. from running the Yukawa couplings up to the GUT scale,
and then running the dimension-5 operators generated by integrating
out the color-triplet Higgs, back down to the weak scale. We have
checked that the effect of the messenger sector fields on the
numerical value of $A_S$ is negligible, and set $A_S \approx 0.67$.
Note that this is an extremely conservative assumption.  The effect
of going to $n_5=2$, for example, reduces $A_S$ to $0.63$.  However,
we have not included the effect of the top quark Yukawa coupling in
the running which can enhance $A_S$ (and strengthen the resulting
proton decay bound) by as much as a factor of $\sim 3$ \cite{hmm}.
We simply set $A_S=0.67$ in our numerical analysis as a conservative
estimate.  $f$ is the ``triangle" function obtained by dressing the
dimension-5 supersymmetric operators with winos to produce
dimension-6 four-fermion operators that are responsible for the
decay \cite{SYW,NA}.  It is given by
\beq
f(u,d) \equiv \frac{m_{\tilde{w}}}{m^2_{\tilde{u}}-m^2_{\tilde{d}}}
\left(\frac{m_{\tilde{u}}}{m^2_{\tilde{u}}-m^2_{\tilde{w}}}
\ln \frac{m^2_{\tilde{u}}}{m^2_{\tilde{w}}} -
\frac{m_{\tilde{d}}}{m^2_{\tilde{d}}-m^2_{\tilde{w}}}
\ln \frac{m^2_{\tilde{d}}}{m^2_{\tilde{w}}} \right) .
\label{eq:trifunc}
\eeq
The squark, slepton and wino masses fed into the triangle function are
those given by the generalization of (\ref{eq:mgaugino}) and
(\ref{eq:msquark}) to the case of $n_5$ {\bf 5}+\bfbar{5} pairs or
$n_{10}$ {\bf 10}+\bfbar{10} pairs, as we
discussed earlier.  The precise values that we choose depend on the value
of the parameter $m_0$, which is of order the messenger scale.  We
choose $m_0$ such that the squark masses are normalized 1 TeV at the
weak scale, and the gaugino masses are fixed by (\ref{eq:mgaugino}).  We
make this choice because the proton-decay bounds are weakest for the
heaviest squark masses. However, we do not allow squark masses greater
than 1 TeV in the interest of naturalness. Finally, the
quantity $1+y^{tK}$ parametrizes possible interference between
the diagrams involving $\tilde{c}$ and $\tilde{t}$ exchange \cite{NA}.
While the bound from the $K\nu$ mode can be made arbitrarily weak
by assuming destructive interference between diagrams (i.e.
$y^{tK}=-1.0$), one cannot simultaneously weaken the bounds from
the other possible decay modes \cite{hmm}.  For example, for $|1+y_{tK}|<0.4$,
the mode $n\rightarrow \pi^0 \overline{\nu}_\mu$ becomes dominant, giving
a comparable decay rate.  Thus, we use the $K\nu$ mode with
$|1+y_{tK}|=0.4$ to obtain a characteristic bound on the parameter space.

With all the parameters fixed as described above, we obtain a bound
on the quantity $\sin 2 \beta$, or alternately $\tan\beta$, the
ratio of up- to down-type Higgs vevs, for each value of $\alpha_3(m_Z)$
that we input in step {\em ii}.  The results are shown in
Figures~2a through 2c.  The area of the $\alpha_3$-$\tan\beta$ plane
that is consistent with the proton decay constraints is the region
above the solid line labeled proton decay.  The region between
the horizontal dashed lines show the area allowed by the two
standard deviation uncertainty in the experimental value of
$\alpha_3(m_Z)$ measured at LEP $0.116\pm 0.005$ \cite{ip}.
It is clear that in going from $n_5=n_{10}=0$ to $n_5=1$, these
regions no longer overlap.  In the $n_5=n_{10}=0$ case, we have chosen
the most conservative set of parameters possible, namely, we have set
all the squark and slepton masses $\tilde{m}=1$ TeV, and the wino
mass $m_{\tilde{w}}=45$ GeV.  However, in the $n_5=1$ case, with squark
masses normalized to 1 TeV, the slepton and gaugino masses are predicted
from (\ref{eq:mgaugino}) and (\ref{eq:msquark}).  Because we have
greater predictivity in this case, we can no longer choose the
most favorable parameter set, and the bound is strengthened.
This effect is large enough to overwhelm the competing effect of the
slightly weaker bound on the color triplet Higgs mass that we
found in Figure~1.  As we pointed out earlier, the ratio of gaugino
to squark mass scales at $\sqrt{n_5}$, so as we go to larger values
of $n_5$, we obtain an even less favorable set of parameters.
Our proton decay bound stays more or less the same however,  due to the
competition between this effect and the weaker bound
on the color triplet Higgs mass.  Finally, we obtain a slightly tighter
bound in the {\bf 10}+\bfbar{10} case, Figure~2c, as a result of the
slightly tighter upper bound on $M_{H_c}$ shown in Figure~1.
For each case in which a messenger sector is present, the proton
decay region never intersects with the allowed region for
$\alpha_3(m_Z)$.

\section{$b$-$\tau$ unification}

For completeness, we have also determined the region of the
$\alpha_3$-$\tan\beta$ plane that is consistent with
$b$-$\tau$ unification.  Here we work only to the one-loop
level.  In our proton-decay analysis, the bound on $M_{H_C}$
followed from threshold corrections, so we needed to include
all other effects of equal importance.  This necessitated the
two-loop analysis.  In the case of $b$-$\tau$ unification,
however, we do not need the higher level of accuracy as we will explain below,
so we worked at one loop.

Our algorithm is straightforward.  For a given choice of $\alpha_3$
and $\tan\beta$, we determine the top, bottom, and tau Yukawa couplings.
We run these Yukawa couplings up to the scale $m_{top}$ using the standard
model renormalization group equations.  Above $m_{top}$ we run these Yukawa
couplings using the one-loop supersymmetric renormalization group equations
up to the GUT scale \cite{bbo}.  We impose two conditions to determine whether
we have a valid solution with $b$-$\tau$ unification: (a) we
require $\lambda_b$ and $\lambda_\tau$ to be within 0.5\% of each other at
the GUT scale (b) we require $\lambda_b$, $\lambda_\tau$
and  $\lambda_{top}$ to be less than 2.   Condition (b) is imposed
so that the couplings do not blow up below 10 times the GUT scale.
If the cutoff scale is lower than this, higer-dimension operators
can generate corrections to the Yukawa unification at more than the 10\% level
\cite{BMR}.  The messenger sector particles alter the
analysis through their effect on the running of the gauge couplings,
which in turn enter into the one-loop RGEs for the Yukawa couplings.
These renormalization group equations are provided in the appendix.
The requirement of unification to within 0.5\% at the GUT scale is
somewhat arbitrary.  Threshold corrections and the effects of
$M_{\rm Planck}$ supressed operators could in principle account
for a larger mismatch between the Yukawa couplings at the GUT scale.
Our results should therefore be considered qualitative.  Unlike
our proton decay analysis, we do not do a two-loop analysis including
threshold corrections at each of the relevant scales.
The presence of $M_{\rm Planck}$ supressed operators can give
important GUT scale threshold corrections in the small $\tan \beta$ region
\cite{BMR}, rendering the extra accuracy of such an analysis meaningless.

Our results are shown in Figures~2a through 2c. The crescent shaped
region is generated by allowing the $b$-quark $\overline{{\rm MS}}$ mass
to vary between 4.1 and 4.5 GeV, the range suggested by the QCD sum rule
analysis \cite{pdg}. Note that as we increase $n_5$ or $n_{10}$, the
$b$-$\tau$ region moves towards smaller values of $\alpha_3$.  While this
is not inconsistent with the measured value of $\alpha_3$, the distance
between the $b$-$\tau$ region and the area allowed by the proton decay
bounds increases monotonically with $n_5$ or $n_{10}$.\footnote{Note that
the low $\tan\beta$ `cusp' of the crescent would extend into the proton
decay region for allowed $\alpha_3(m_Z)$ in the minimal
case ($n_5 = n_{10} = 0$) had we imposed the less stringent constraint
$\lambda_{top}<3.3$ \cite{bbo} on the acceptable size of the Yukawa
couplings at the GUT scale.  Even in this case, however, the three-way
overlap between proton decay, $b$-$\tau$ unification and $\alpha_3(m_Z)$,
does not persist when the messenger fields are present.  If we had chosen
a smaller value of $m_{top}=150$ GeV, on the other hand, the $b$-$\tau$
region would extend to smaller $\tan\beta$, but not to larger
$\alpha_3(m_Z)$, and our conclusions would remain the same.}

\section{Conclusions}

We have shown that the minimal supersymmetric SU(5) model
cannot be embedded successfully within the simplest type of scenario
suggested in Ref.~\cite{nd2}.  The effect of messenger sector particles
on the renormalization group analysis, as well as the predicted
gaugino and squark mass ratios that follow from the messenger sector
particle content lead to a conflict with the lower bound on the proton
lifetime.  In addition, the region of parameter space preferred
for $b$-$\tau$ unification moves farther away from the region preferred
by the proton decay bounds as the size of the messenger sector is
increased.  We must emphasize that this is only a mild obstacle
to the type of scenario proposed in Ref.~\cite{nd2}.  Non-minimal GUTs
can be constructed which evade the proton decay bound \cite{babu}.
It is nonetheless interesting that we can exclude what is perhaps
the {\em simplest} models of this type.
\begin{center}
{\bf Acknowledgments}
\end{center}

This work was supported in part by the Director, Office of
Energy Research, Office of High Energy and Nuclear Physics, Division of
High Energy Physics of the U.S. Department of Energy under Contract
DE-AC03-76SF00098 and in part by the National Science Foundation under
grant PHY-90-21139.

\appendix
\section{Appendix}

The two-loop supersymmetric renormalization group equation for the
gauge couplings that we use in our proton decay analysis is
\beq
\mu \frac{\partial g_i}{\partial \mu}= \frac{1}{16\pi^2} b_i g_i^3
+\left(\frac{1}{16\pi^2}\right)^2 \sum_{j=1}^3 b_{ij} g_i^3 g_j^2
\eeq
where the beta functions are given by
\beq
b_i =
\left[\begin{array}{c} 2 \\ 2 \\ 2 \end{array}\right] n_g +
\left[\begin{array}{c} \frac{3}{10} \\ \frac{1}{2} \\ 0
\end{array}\right] n_h +
\left[\begin{array}{c} 1 \\ 1 \\ 1 \end{array}\right] n_5 +
\left[\begin{array}{c} 3 \\ 3 \\ 3 \end{array}\right] n_{10}+
\left[\begin{array}{c} 0 \\ -6 \\ -9 \end{array}\right]
\eeq
\[
b_{ij}=
\left(\begin{array}{ccc} \frac{38}{15} & \frac{6}{5} & \frac{88}{15} \\
\frac{2}{5} & 14 & 8 \\ \frac{11}{15} & 3 & \frac{68}{3} \end{array}
\right) n_g +
\left(\begin{array}{ccc} \frac{9}{50} & \frac{9}{10} & 0 \\
\frac{3}{10} & \frac{7}{2} & 0 \\ 0 & 0 & 0 \end{array}
\right) n_h
\]
\[
\,\,\,+\left(\begin{array}{ccc} \frac{21}{45} & \frac{9}{5} & \frac{32}{15} \\
\frac{3}{5} & 7 & 0 \\ \frac{4}{15} & 0 & \frac{34}{3} \end{array}
\right) n_5
+\left(\begin{array}{ccc} \frac{23}{5} & \frac{3}{5} & \frac{48}{5} \\
\frac{1}{5} & 21 & 16 \\ \frac{6}{5} & 6 & 34 \end{array}
\right) n_{10}
\]
\beq
+\left(\begin{array}{ccc} 0 & 0 & 0 \\
0 & -24 & 0 \\ 0 & 0 & -54 \end{array}
\right) \,\, .
\eeq
Here $n_g$, $n_h$, $n_5$ and $n_{10}$ are the number of generations (3),
higgs doublets (2), messenger sector {\bf 5}+\bfbar{5} and
{\bf 10}+\bfbar{10} pairs, respectively.

The one-loop renormalization group equations for the top, bottom,
and tau Yukawa couplings used in our analysis of $b$-$\tau$
Yukawa unification are given by \cite{bbo}
\beq
\mu \frac{\partial \lambda_{top}}{\partial \mu} =
\frac{1}{16\pi^2}\left(-\sum_i c_i g_i^2 + \lambda_b^2 +
6 \lambda_{top}^2\right)
\lambda_{top} \,\, ,
\eeq
\beq
\mu \frac{\partial \lambda_{b}}{\partial \mu} =
\frac{1}{16\pi^2}\left(-\sum_i c'_i g_i^2 + \lambda_\tau^2
+ 6 \lambda_b^2 +
\lambda_{top}^2\right)
\lambda_b \,\, ,
\eeq
\beq
\mu \frac{\partial \lambda_{\tau}}{\partial \mu} =
\frac{1}{16\pi^2}\left(-\sum_i c''_i g_i^2 +
4 \lambda_\tau^2+3\lambda_b^2 \right)
\lambda_\tau \,\, ,
\eeq
where $c_i=(\frac{13}{15}\mbox{, }3 \mbox{, } \frac{16}{3})$,
$c'_i=(\frac{7}{15}\mbox{, }3 \mbox{, } \frac{16}{3})$,
and $c''_i=(\frac{9}{5}\mbox{, }3 \mbox{, } 0)$.


\newpage

\begin{center}
{\bf Figure Captions}
\end{center}

{\bf Fig. 1} (a) Upper bound on the color-triplet Higgs mass $M_{H_C}$
as a function of $\alpha_3(m_Z)$, ignoring threshold corrections
(i.e. mass splittings) at the messenger scale.  The solid line is the minimal
SU(5) result, the dashed line is the case of three {\bf 5}+\bfbar{5}
pairs in the messenger sector, and the dotted line is the case of
one {\bf 10}+\bfbar{10} pair.  (b) The complete result, including
threshold corrections at the messenger scale.

{\bf Fig. 2}(a) Preferred regions of the $\tan(\beta)$-$\alpha_3(m_Z)$
plane, in minimal SU(5) unification, for $m_{top}=176$ GeV.  The region
above the proton-decay line is allowed by the lower bound on the proton
lifetime (see the text) at 90\% confidence level,
while the region within the crescent shaped contour is preferred by the
constraint of $b$-$\tau$ Yukawa unification with $m_b(m_b) = 4.1$--$4.5$~GeV.
The horizontal band shows the experimentally allowed range
of $\alpha_3(m_Z) = 0.116\pm0.005$ at two standard deviations. (b) Same
as (a) for $n_5=1$, (c) Same as (a) for $n_5=3$.
The dashed line shows the result for $n_{10}=1$ and $n_5=0$; the $b$-$\tau$
region remains the same for $n_5=3$ or $n_{10}=1$.

\begin{thebibliography}{99}
\frenchspacing
\bibitem{sur}
For a review, see  R.~Mohapatra, {\em Unification and Supersymmetry},
(Springer-Verlag, Berlin, 1992); G.G.~Ross, {\em Grand Unified Theories},
(Addison-Wesley, Reading, MA, 1984).
\bibitem{msr}
M. Veltman, {\sl Acta Phys. Pol.}\/ {\bf B12}, 437 (1981);
L. Maiani, in {\it Proceedings of the Eleventh Gif-sur-Yvette Summer
School on Particle Physics}, Gif-sur-Yvette, France, 1979 ({\sl Inst. Nat.
Phys. Nucl. Phys. Particules},\/ Paris, 1980), p.3;
S. Dimopoulos and S. Raby {\sl Nucl. Phys.}\/ {\bf B192}, 353 (1981);
E. Witten, {\sl Nucl. Phys.}\/ {\bf B188}, 513 (1981);
M. Dine, W. Fischler and M. Srednicki, {\sl Nucl. Phys.}\/ {\bf B189}, 575
(1981).
\bibitem{ymsr}
U.~Amaldi, W.~de Boer, and H.~Furstenau, {\sl Phys. Lett. }\/
{\bf B260} 447 (1991); J.~Ellis, S.~Kelley, and D.V.~Nanopoulos,
{\sl Phys. Lett. }\/{\bf B260} 131 (1991); P.~Langacker and
M.~Luo, {\sl Phys. Rev.} \/ {\bf D44}, 817 (1991).
\bibitem{nd1}
M.~Dine and A.E.~Nelson, {\sl Phys. Rev.}\/ {\bf D48},
1277 (1993).
\bibitem{nd2}
M.~Dine, A.E.~Nelson and Y.~Shirman, {\sl Phys. Rev.}\/ {\bf D51},
1362 (1995);
M.~Dine, A.E.~Nelson, Y.~Nir and Y.~Shirman,
   SCIPP-95-32 (unpublished), hep-ph/9507378.
\bibitem{DGS}
S.~Dimopoulos and H.~Georgi, {\sl Nucl. Phys.}\/ {\bf B193}, 150 (1981);
N.~Sakai, {\sl Z. Phys.}\/ {\bf C11}, 153 (1981).
\bibitem{wise}
L.~Alvarez-Gaum\'e, M.~Claudson and M.B.~Wise, {\sl Nucl. Phys.} \/
{\bf B207}, 96 (1982).
\bibitem{ads}
I.~Affleck, M.~Dine, and N.~Seiberg, {\sl Nucl. Phys.} \/
{\bf B256}, 557 (1985).
\bibitem{gfarrar}
See H.E.~Haber, in Proceedings of the International Workshop on
Supersymmetry and Unification of Fundamental Interactions (SUSY 93),
Edited by Pran Nath. River Edge, N.J., World Scientific, 1993, p. 373.
\bibitem{MMY}
T. Moroi, H. Murayama, and T. Yanagida, {\sl Phys. Rev.}\/ {\bf D48}, 2995
(1993);
B. Brahmachari, U. Sarkar, and K. Sridhar,
{\sl Mod. Phys. Lett.}\/ {\bf A8}, 3349 (1993).
\bibitem{hmm}
J.~Hisano, H.~Murayama, and T.~Yanagida, {\sl Nucl. Phys.}\/ {\bf
B402}, 46 (1993).
\bibitem{HMTY} J.~Hisano, T.~Moroi, K.~Tobe, and T.~Yanagida,
TU-470 (unpublished), hep-ph/9411298.
\bibitem{ip}
J.~Erler and P.~Langacker, {\sl Phys. Rev.}\/ {\bf D52}, 441 (1995).
\bibitem{ENR}
J.~Ellis, D.V.~Nanopoulos and S.~Rudaz, {\sl Nucl. Phys.}\/ {\bf B202}, 43
(1982).
\bibitem{SYW}
N.~Sakai and T.~Yanagida, {\sl Nucl. Phys.}\/ {\bf B197}, 533 (1982);
S.~Weinberg, {\sl Phys. Rev.}\/ {\bf D26}, 287 (1982).
\bibitem{NA}
P.~Nath, A.H.~Chamseddine and R.~Arnowitt, {\sl Phys. Rev.}\/ {\bf D32}, 2348
(1985);
P.~Nath and R.~Arnowitt, {\sl Phys. Rev.}\/ {\bf D38}, 1479 (1988).
\bibitem{BMR}
A.~Brignole, H.~Murayama and R.~Rattazzi,
{\sl Phys. Lett.}\/ {\bf B335}, 345 (1994).
\bibitem{bbo}
V.~Barger, M.S.~Berger, and P.~Ohmann, {\sl Phys. Rev.}\/ {\bf D47},
1093 (1993).
\bibitem{pdg}
{\em Review of Particle Properties}\/,
Particle Data Group (L.~Montanet {\it et al.}\/),
{\sl Phys. Rev.} \/{\bf D50},
1173 (1994).
\bibitem{babu}
K.S.~Babu and S.M.~Barr, {\sl Phys. Rev.}\/ {\bf D48},
5354 (1993).
\end{thebibliography}
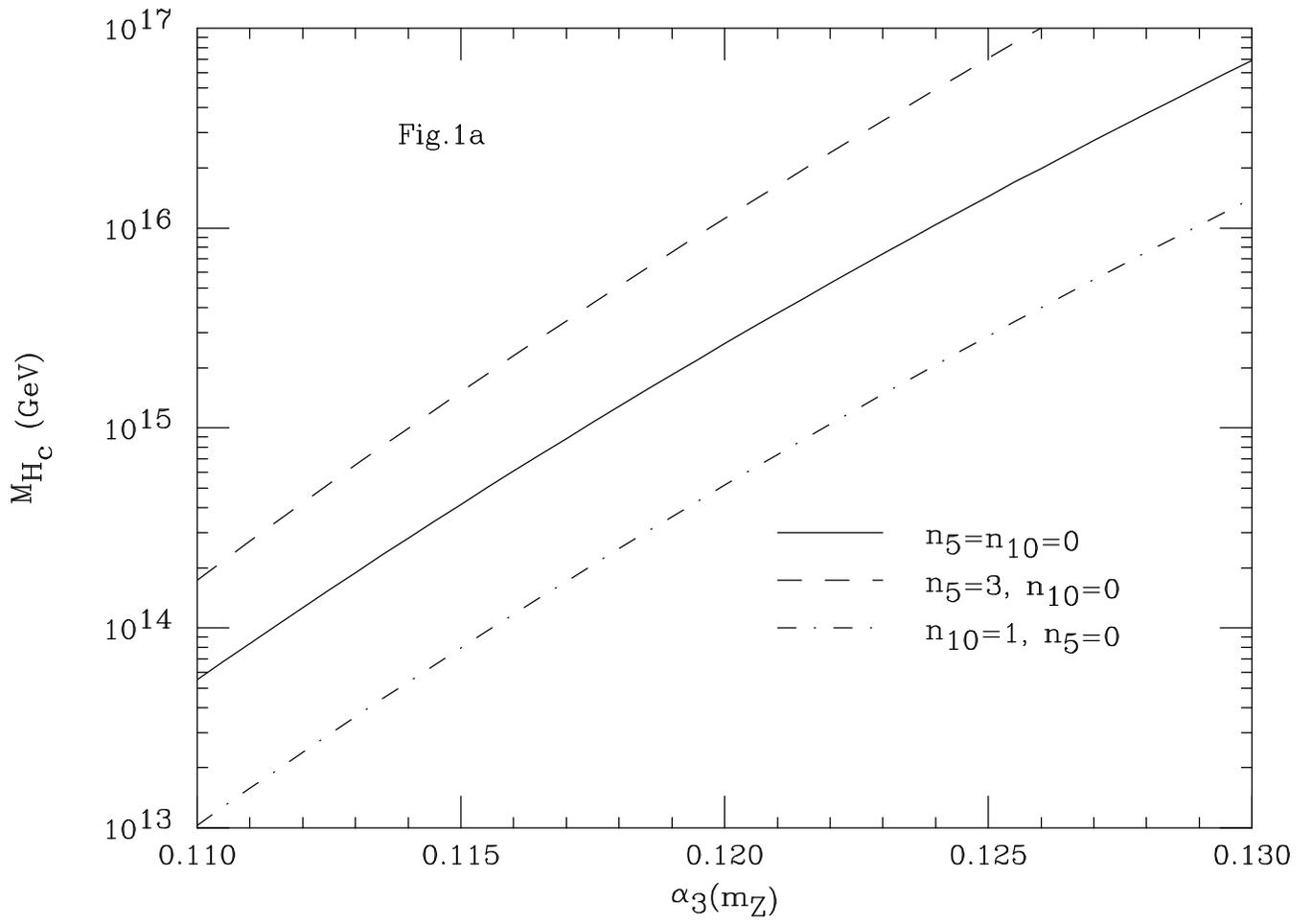
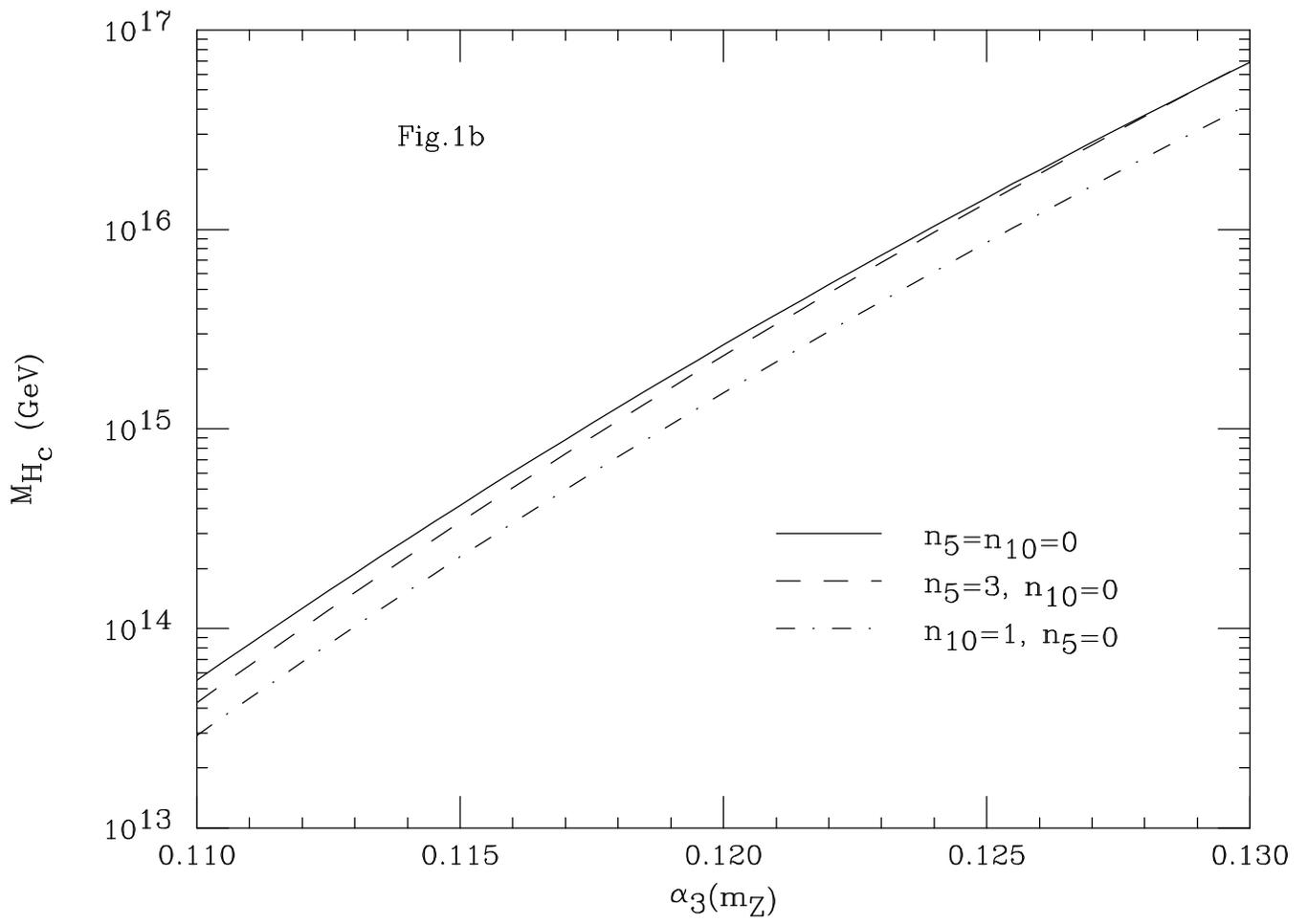
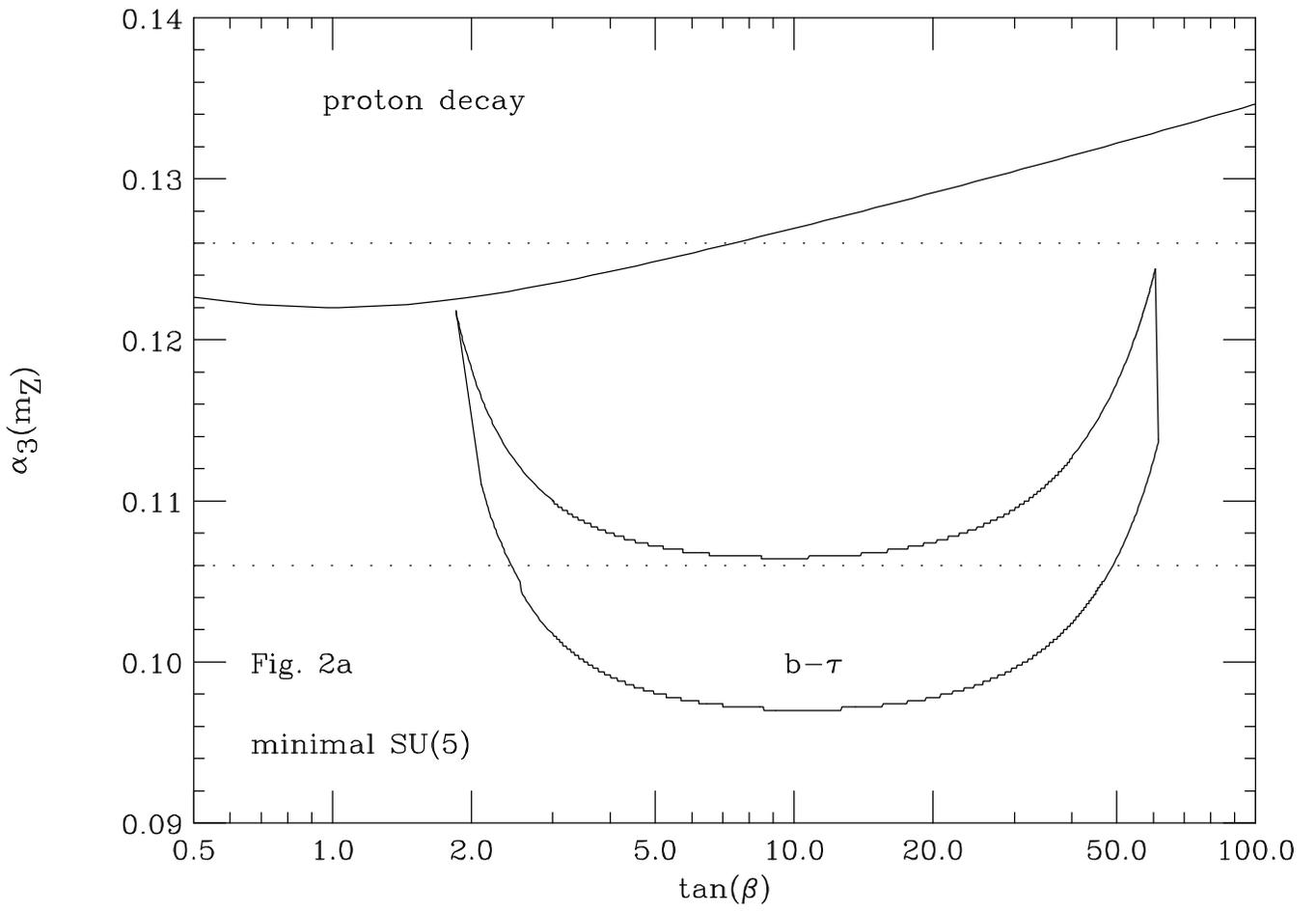
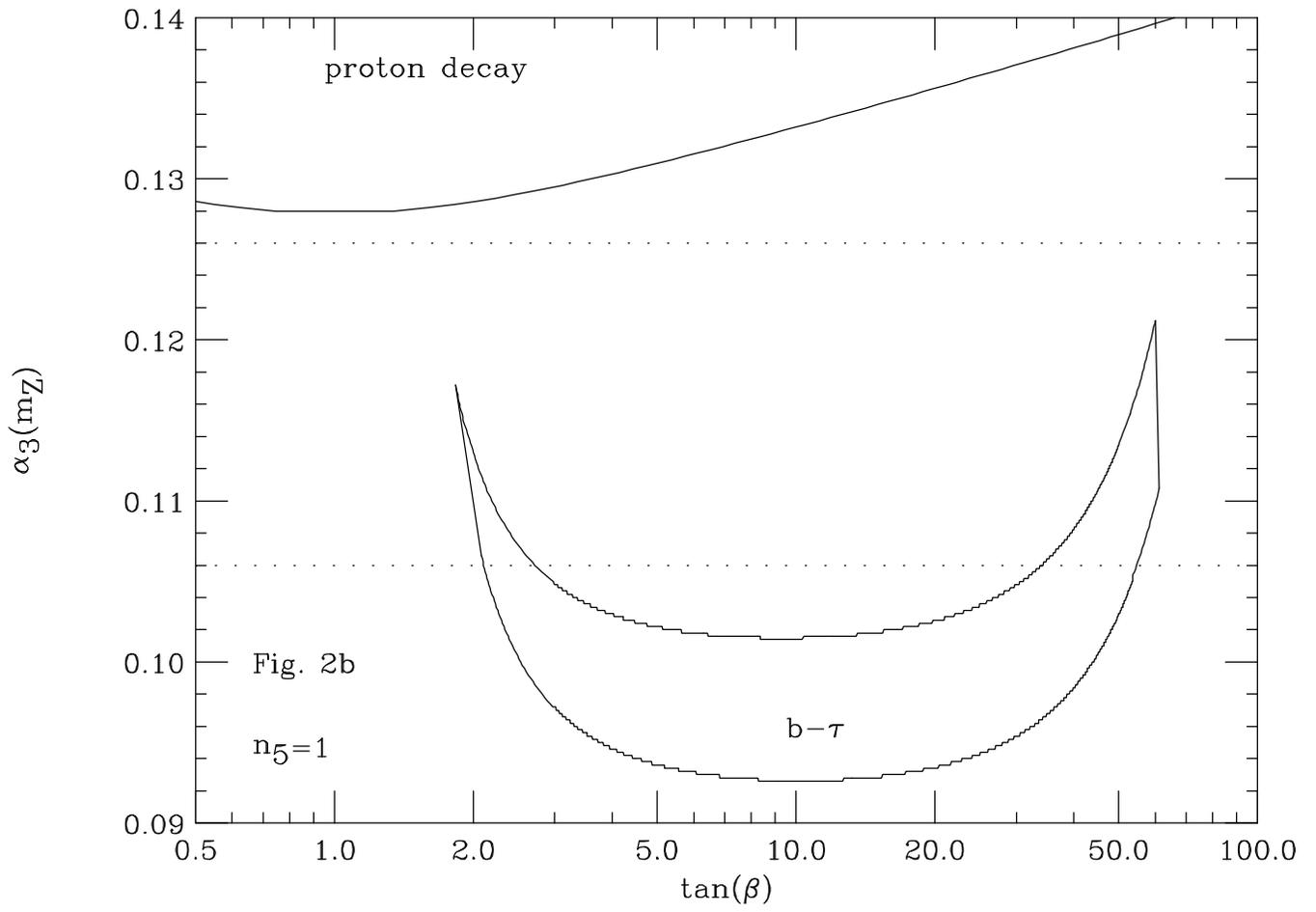
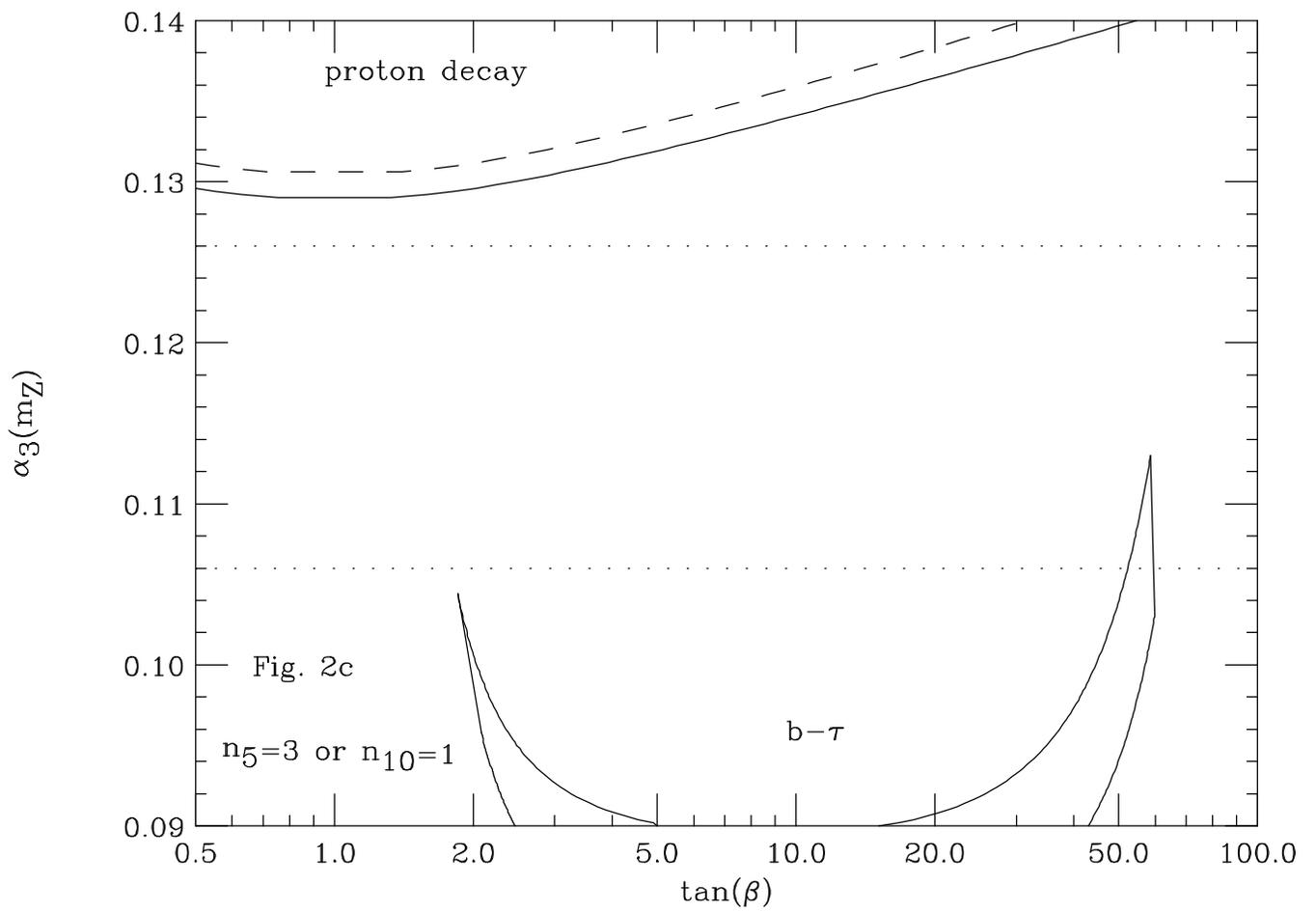
\end{document}